\documentclass[11pt,a4paper]{article}
\usepackage{drdo_iisc}

\usepackage{subfigure}
\usepackage{setspace}
\usepackage{amsmath}
\usepackage{amssymb}
\usepackage{amsfonts}
\usepackage{amscd}
\usepackage{mathrsfs}
\usepackage{graphicx}
\usepackage{color}

\newtheorem{eg}{Example}

\newtheorem{note}{Note}

\begin{document}
\def\proof{\noindent\hspace{0em}{\itshape \textbf{Proof} :}}
\def\appendices{\noindent\hspace{0em}{\itshape Appendices: }}
\def\QED{\mbox{\rule[0pt]{1.3ex}{1.3ex}}}
\def\endproof{\hspace*{\fill}~\QED\par\endtrivlist\unskip}

\title{Distributed Space Time Codes with Low Decoding Complexity for Asynchronous Relay Networks}
\author{G. Susinder Rajan and B. Sundar Rajan\footnotemark[1]}

\affiliation{Department of ECE, \\Indian Institute of Science, Bangalore}
\reportnumber{TR-PME-2007-09}
\date{\today}

\titlepage
\maketitle

\footnotetext[1]{The authors are with the Department of Electrical Communication Engineering, Indian Institute of Science, Bangalore- 560012, India, email:\{susinder,bsrajan\}@ece.iisc.ernet.in\\
This work was partly supported by the DRDO-IISc Program of Advanced Research in Mathematical Engineering through a grant to B.S.Rajan}

\begin{abstract}
Recently Li and Xia have proposed a transmission scheme for wireless relay networks based on the Alamouti space time code and orthogonal frequency division multiplexing to combat the effect of timing errors at the relay nodes. This transmission scheme is amazingly simple and achieves a diversity order of two for any number of relays. Motivated by its simplicity, this scheme is extended to a more general transmission scheme that can achieve full cooperative diversity for any number of relays. The conditions on the distributed space time code (DSTC) structure that admit its application in the proposed transmission scheme are identified and it is pointed out that the recently proposed full diversity four group decodable DSTCs from precoded co-ordinate interleaved orthogonal designs and extended Clifford algebras satisfy these conditions. It is then shown how differential encoding at the source can be combined with the proposed transmission scheme to arrive at a new transmission scheme that can achieve full cooperative diversity in asynchronous wireless relay networks with no channel information and also no timing error knowledge at the destination node. Finally, four group decodable distributed differential space time codes applicable in this new transmission scheme for power of two number of relays are also provided. 
\end{abstract}

\section{Introduction}
\label{sec1}

Coding for cooperative wireless relay networks has attracted considerable attention recently. Distributed space time coding was proposed as a coding strategy to achieve full cooperative diversity in \cite{JiH} assuming that the signals from all the relay nodes arrive at the destination at the same time. But this assumption is not close to practicality since the relay nodes are geographically distributed. In \cite{GuX}, a transmission scheme based on orthogonal frequency division multiplexing (OFDM) at the relay nodes was proposed to combat the timing errors at the relays and a high rate space time code (STC) construction was also provided. However, the maximum likelihood (ML) decoding complexity for this scheme is prohibitively high especially for the case of large number of relays. Several other works in the literature propose methods to combat the timing offsets but most of them are based on decode and forward at the relay node and moreover fail to address the decoding complexity issue. In \cite{LiX}, a simple transmission scheme to combat timing errors at the relay nodes was proposed. This scheme is particularly interesting because of its associated low ML decoding complexity. In this scheme, OFDM is implemented at the source node and time reversal/conjugation is performed at the relay nodes on the received OFDM symbols from the source node. The received signals at the destination after OFDM demodulation are shown to have the Alamouti code structure and hence single symbol maximum likelihood (ML) decoding can be performed. However, the Alamouti code is applicable only for the case of two relay nodes and for larger number of relays, the authors of \cite{LiX} propose to cluster the relay nodes and employ Alamouti code in each cluster. But this clustering technique provides diversity order of only two and fails to exploit the full cooperative diversity equal to the number of relay nodes.     

The main contributions of this report are as follows.
\begin{itemize}
\item The Li-Xia transmission scheme is extended to a more general transmission scheme that can achieve full asynchronous cooperative diversity for any number of relays.
\item The conditions on the STC structure that admit its application in the proposed transmission scheme are identified. The recently proposed full diversity four group decodable distributed STCs in \cite{RaR1,RaR2,RaR3} for synchronous wireless relay networks are found to satisfy the required conditions for application in the proposed transmission scheme.
\item It is shown how differential encoding at the source node can be combined with the proposed transmission scheme to arrive at a transmission scheme that can achieve full asynchronous cooperative diversity in the absence of channel knowledge and in the absence of knowledge of the timing errors of the relay nodes. Moreover, an existing class of four group decodable distributed differential STCs \cite{RaR4} for synchronous relay networks with power of two number of relays is shown to be applicable in this setting as well.  
\end{itemize}

\subsection{Organization of the report}
\label{subsec1_1}
In Section \ref{sec2}, the basic assumptions on the relay network model are given and the Li-Xia transmission scheme is briefly described. Section \ref{sec3} describes the transmission scheme proposed in this report and also provides four group decodable codes for any number of relays. Section \ref{sec4} briefly explains how differential encoding at the source node can be combined with the proposed transmission scheme and four group decodable distributed differential STCs applicable in this scenario are also proposed. Simulation results and discussion on further work comprise Sections \ref{sec5} and \ref{sec6} respectively.\\
~\\ 
\noindent
\textbf{Notation:}
Vectors and matrices are denoted by lowercase and uppercase bold letters respectively. $\mathbf{I_m}$ denotes an $m\times m$ identity matrix and $\mathbf{0}$ denotes an all zero matrix of appropriate size. For a set $A$, the cardinality of $A$ is denoted by $|A|$. A null set is denoted by $\phi$. For a matrix, $(.)^T$, $(.)^*$ and $(.)^H$ denote transposition, conjugation and conjugate transpose operations respectively. For a complex number, $(.)_I$ and $(.)_Q$ denote its in-phase and quadrature-phase parts respectively. 

\section{Relay network model assumptions and the Li-Xia transmission scheme}
\label{sec2}

In this section, the basic relay network model assumptions are given and the Li-Xia transmission scheme in \cite{LiX} is briefly described. The transmission scheme in \cite{LiX} is based on the use of OFDM at the source node and the Alamouti code implemented in a distributed fashion for a $2$ relay system. Essentially, the transmission scheme in \cite{LiX} is applicable mainly for the case of $2$ relays but by forming clusters of two relay nodes, it can be extended to more number of relays at the cost of sacrificing diversity benefits. 

\begin{figure}[h] 
\centering 
\input{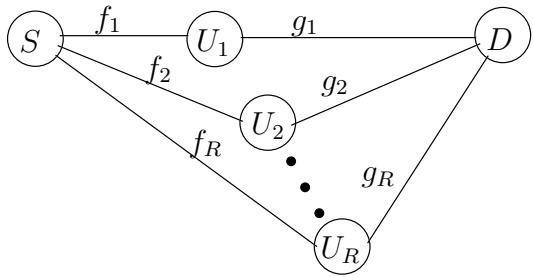} 
\caption{Asynchronous wireless relay network}
\label{fig_network}
\end{figure}

\subsection{Network model assumptions}
\label{subsec2_1}
Consider a network with one source node, one destination node and $R$ relay nodes $U_1, U_2, \dots, U_R$. This is depicted in Fig. \ref{fig_network}. Every node is assumed to have only a single antenna and is half duplex constrained.  The channel gain between the source and the $i$-th relay $f_i$ and that between the $j$-th relay and the destination $g_j$ are assumed to be quasi-static, flat fading and modeled by independent and complex Gaussian distributed with mean zero and unit variance. The transmission of information from the source node to the destination node takes place in two phases. In the first phase, the source broadcasts the information to the relay nodes using OFDM. The relay nodes receive the faded and noise corrupted OFDM symbols, process them and transmit them to the destination. The relay nodes are assumed to have perfect carrier synchronization. The overall relative timing error of the signals arrived at the destination node from the $i$-th relay node is denoted by $\tau_i$. Without loss of generality, it is assumed that $\tau_1=0$, $\tau_{i+1}\geq \tau_i,i=1,\dots,R-1$. The destination node is assumed to have the knowledge of all the channel fading gains $f_i,g_j, i,j=1,\dots,R$ and the relative timing errors $\tau_i,i=1,\dots,R$.

\subsection{Li-Xia transmission scheme\cite{LiX}}
\label{subsec2_2}

The source takes $2N$ complex symbols $x_{i,j, 0\leq i\leq N-1, j=1,2}$ and forms two blocks of data denoted by $\mathbf{x_j}=\left[\begin{array}{cccc}x_{0,j} & x_{1,j} & \dots & x_{N-1,j}\end{array}\right]^T, j=1,2$. The first block $\mathbf{x_1}$ is modulated by $N$-point Inverse Discrete Fourier Transform (IDFT) and $\mathbf{x_2}$ is modulated by $N$-point Discrete Fourier Transform (DFT). Then a cyclic prefix (CP) of length $l_{cp}$ is added to each block, where $l_{cp}$ is not less than the maximum of the overall relative timing errors of the signals arrived at the destination node from the relay nodes. The resulting two OFDM symbols denoted by $\mathbf{\bar{x}_1}$ and $\mathbf{\bar{x}_2}$ consisting of $L_s=N+l_{cp}$ complex numbers are broadcasted to the two relays using a fraction $\pi_1$ of the total average $P$ consumed by the source and the relay nodes together. 

If the channel fade gains are assumed to be constant for $4$ OFDM symbol intervals, the received signals at the $i$-th relay during the $j$-th OFDM symbol duration is given by 

$$
\mathbf{r_{i,j}}=f_i\mathbf{\bar{x}_j}+\mathbf{\bar{v}_{i,j}}
$$

\noindent where, $\mathbf{\bar{v}_{i,j}}$ is the additive white Gaussian noise at the $i$-th relay node during the $j$ the OFDM symbol duration. The two relay nodes then process and transmit the resulting signals as shown in Table \ref{table_alamouti} using a fraction $\pi_2$ of the total power $P$. The notation $\zeta(.)$ denotes the time reversal operation, i.e., $\zeta(\mathbf{r}(n))\triangleq \mathbf{r}(L_s-n)$.

\begin{table}[h]
\caption{Alamouti code based transmission scheme}
\label{table_alamouti}
\begin{center}
\begin{tabular}{|c|c|c|}
\hline
OFDM Symbol & $U_1$ & $U_2$\\
\hline
$1$ & $\sqrt{\frac{\pi_2P}{\pi_1P+1}}\mathbf{r_{1,1}}$ & $-\sqrt{\frac{\pi_2P}{\pi_1P+1}}\mathbf{r_{2,2}}^*$\\
\hline
$2$ & $\sqrt{\frac{\pi_2P}{\pi_1P+1}}\zeta(\mathbf{r_{1,2}})$ & $\sqrt{\frac{\pi_2P}{\pi_1P+1}}\zeta(\mathbf{r_{2,1}}^*)$\\
\hline
\end{tabular}
\end{center}
\end{table}

The destination removes the CP for the first OFDM symbol and implements the following for the second OFDM symbol:

\begin{enumerate}
\item Remove the CP to get a $N$-point vector 
\item Shift the last $l_{cp}$ samples of the $N$-point vector as the first $l_{cp}$ samples.
\end{enumerate} 

DFT is then applied on the resulting two vectors. Since $l_{cp}\geq\tau_2$, the orthogonality between the sub carriers is still maintained. The delay in the time domain then translates to a corresponding phase change of $e^{-\frac{i2\pi k}{N}}$ in the $k$-th sub carrier. Let $d^{\tau_2}$ denote $\left[\begin{array}{cccc}1 & e^{-\frac{i2\pi\tau_2}{N}} & \dots & e^{-\frac{i2\pi\tau_2(N-1)}{N}} \end{array}\right]^T$. Then the received signals for two consecutive OFDM blocks after CP removal and DFT transformation denoted by $\mathbf{y_1}=\left[\begin{array}{cccc}y_{0,1} & y_{1,1} & \dots & y_{N-1,1}\end{array}\right]^T$ and $\mathbf{y_2}=\left[\begin{array}{cccc}y_{0,2} & y_{1,2} & \dots & y_{N-1,2}\end{array}\right]^T$ can be expressed as:  

$$
\begin{array}{rl}
\mathbf{y_1}=& \sqrt{\frac{\pi_1\pi_2P^2}{\pi_1P+1}}(\mathrm{DFT}(\mathrm{IDFT}(\mathbf{x_1}))f_1g_1
+\mathrm{DFT}(-(\mathrm{DFT}(\mathbf{x_2}))^*)\circ d^{\tau_2}f_2^*g_2)\\
&+\sqrt{\frac{\pi_2P}{\pi_1P+1}}(\mathbf{v_{1,1}}g_1-\mathbf{v_{2,1}}^*\circ d^{\tau_2}g_2)+\mathbf{w_1}\\
\mathbf{y_2}=& \sqrt{\frac{\pi_1\pi_2P^2}{\pi_1P+1}}(\mathrm{DFT}(\zeta(\mathrm{DFT}(\mathbf{x_2}))^*)f_1g_1
+\mathrm{DFT}(\zeta(\mathrm{IDFT}(\mathbf{x_1}))^*)\circ d^{\tau_2}f_2^*g_2)\\
&+\sqrt{\frac{\pi_2P}{\pi_1P+1}}(\mathbf{v_{1,2}}g_1+\mathbf{v_{2,2}}^*\circ d^{\tau_2}g_2)+\mathbf{w_2}
\end{array}
$$

\noindent where, $\circ$ is the Hadamard product, $\mathbf{w_i}=(w_{k,i}), i=1,2$ are the additive white Gaussian noise at the destination and $\mathbf{v_{i,j}}$ denotes the DFT of $\mathbf{\bar{v}_{i,j}}$. Now using the identities 

\begin{equation}
\label{eqn_identities}
(\mathrm{DFT}(\mathbf{x}))^*=\mathrm{IDFT}(\mathbf{x}^*),~ (\mathrm{IDFT}(\mathbf{x}))^*=\mathrm{DFT}(\mathbf{x}^*),~  \mathrm{DFT}(\zeta(\mathrm{DFT}(\mathbf{x})))=\mathbf{x}
\end{equation}

\noindent we get the Alamouti code form in each sub carrier $k,0\leq k\leq N-1$ as shown below: 

$$
\begin{array}{ll}
\left[\begin{array}{c}y_{k,1}\\y_{k,2}\end{array}\right]=&\sqrt{\frac{\pi_1\pi_2P^2}{\pi_1P+1}}\left[\begin{array}{cc}
x_{k,1} & -x_{k,2}^*\\
 x_{k,2} & x_{k,1}^*
\end{array}\right]\left[\begin{array}{c}f_1g_1\\e^{-\frac{i2\pi k\tau_2}{N}}f_2^*g_2\end{array}\right]+\\
&\sqrt{\frac{\pi_2P}{\pi_1P+1}}\left[\begin{array}{c}\mathbf{v_{1,1}}(k)g_1-\mathbf{v_{2,1}}^*(k)e^{-\frac{i2\pi k\tau_2}{N}}g_2\\
\mathbf{v_{1,2}}(k)g_1+\mathbf{v_{2,2}}^*(k)e^{-\frac{i2\pi k\tau_2}{N}}g_2\end{array}\right]+\left[\begin{array}{c}w_{k,1}\\w_{k,2}\end{array}\right].
\end{array}
$$

With the power allocation $\pi_1=1$, $\pi_2=\frac{1}{R}$ and because of the Alamouti code form, diversity order of two can be achieved along with symbol-by-symbol ML decoding.

\section{Proposed Transmission Scheme}
\label{sec3}

In this section, we extend the Li-Xia transmission scheme to a general transmission scheme that can achieve full cooperative diversity for arbitrary number of relays. This nontrivial extension is based on analyzing the sufficient conditions required on the structure of STBCs which admit application in the Li-Xia transmission scheme. 

\subsection{Transmission by the source node}
\label{subsec3_1}

The source takes $RN$ complex symbols $x_{i,j, 0\leq i\leq N-1, j=1,2,\dots,R}$ and forms $R$ blocks of data denoted by $\mathbf{x_j}=\left[\begin{array}{cccc}x_{0,j} & x_{1,j} & \dots & x_{N-1,j}\end{array}\right]^T, j=1,2,\dots,R$. Of these $R$ blocks, $M$ of them are modulated by $N$-point IDFT and the remaining $R-M$ blocks are modulated by $N$-point DFT. Without loss of generality, let us assume that the first $M$ blocks are modulated by $N$-point IDFT. Then a CP of length $l_{cp}$ is added to each block, where $l_{cp}$ is not less than the maximum of the overall relative timing errors of the signals arrived at the destination node from all the relay nodes. The resulting $R$ OFDM symbols denoted by $\mathbf{\bar{x}_1},\mathbf{\bar{x}_2},\dots,\mathbf{\bar{x}_R}$ consisting of $L_s=N+l_{cp}$ complex numbers are broadcasted to the $R$ relays using a fraction $\pi_1$ of the total average $P$.

\subsection{Processing at the relay nodes}
\label{subsec3_2}

If the channel fade gains are assumed to be constant for $2R$ OFDM symbol intervals, the received signals at the $i$-th relay during the $j$-th OFDM symbol duration is given by 

$$
\mathbf{r_{i,j}}=f_i\mathbf{\bar{x}_j}+\mathbf{\bar{v}_{i,j}}
$$

\noindent where, $\mathbf{\bar{v}_{i,j}}$ is the additive white Gaussian noise (AWGN) at the $i$-th relay node during the $j$ the OFDM symbol duration. The relay nodes process and transmit the received noisy signals as shown in Table \ref{table_proposed} using a fraction $\pi_2$ of total power $P$. Note from Table \ref{table_proposed} that time reversal is done during the last $R-M$ OFDM symbol durations. We would like to emphasize that in general time reversal could be implemented in any $R-M$ of the total $R$ OFDM symbol durations. The transmitted signal $\mathbf{t_{i,j}}\in\left\{\mathbf{0},\pm \mathbf{r_{i,j}}, j=1,\dots,R\right\}$ with the constraint that the $i$-th relay should not be allowed to transmit the following: 

$$
\begin{array}{l}
\left\{\pm\mathbf{r_{i,j}}^*, j=1,\dots,M\right\}
\cup\left\{\pm\zeta(\mathbf{r_{i,j}}),j=1,\dots,M\right\}\\
\cup\left\{\pm\mathbf{r_{i,j}},j=M+1,\dots,R\right\}
\cup\left\{\pm\zeta(\mathbf{r_{i,j}}^*), j=M+1,\dots,R\right\}.
\end{array}
$$ 

\begin{note}
If the $i$-th relay is permitted to transmit elements belonging to the above set, then after CP removal and DFT transformation at the destination node, we would end up with the following vectors corresponding to each of the four subsets in the above set respectively:
$$
\begin{array}{l}
\pm\mathrm{DFT}((\mathrm{IDFT}(\mathbf{x_j}))^*)=\mathrm{DFT}(\mathrm{DFT}(\mathbf{x_j}^*)),\ j=1,\dots,M\\
\pm\mathrm{DFT}(\zeta(\mathrm(IDFT)(\mathbf{x_j}))),\ j=1,\dots,M\\
\pm\mathrm{DFT}(\mathrm{DFT}(\mathbf{x_j})),\ j=M+1,\dots,R\\
\pm\mathrm{DFT}(\zeta(\mathrm{DFT}(\mathbf{x_j}))^*)=\pm\mathrm{DFT}(\zeta(\mathrm{IDFT}(\mathbf{x_j}^*))),\ j=M+1,\dots,R
\end{array}
$$

\noindent from any of which it is not possible to recover any of $\pm\mathbf{x_j},\pm\mathbf{x_j}^*,j=1,2,\dots,R$. However, if the destination node is allowed to apply DFT to some of the received OFDM symbols and IDFT to the remaining OFDM symbols, then possibly the above restrictions can be removed, which is a scope for further work.
\end{note}

\begin{table}
\caption{Proposed transmission scheme}
\label{table_proposed}
{\footnotesize
\begin{center}
\begin{tabular}{|c|c|c|c|c|c|c|}
\hline
OFDM Symbol & $U_1$ & $\dots$ & $U_{M}$ & $U_{M+1}$ & $\dots$ & $U_R$\\
\hline
$1$ & $\mathbf{t_{1,1}}$ & $\dots$ & $\mathbf{t_{M,1}}$ & $\mathbf{t_{M+1,1}}^*$ & $\dots$ & $\mathbf{t_{R,1}}^*$\\
\hline
$\vdots$ & $\vdots$ & $\vdots$ & $\vdots$ & $\vdots$ & $\vdots$ & $\vdots$\\
\hline
$M$ & $\mathbf{t_{1,M}}$ & $\dots$ & $\mathbf{t_{M,M}}$ & $\mathbf{t_{M+1,M}}^*$ & $\dots$ & $\mathbf{t_{R,M}}^*$\\
\hline
$M+1$ & $\zeta(\mathbf{t_{1,M+1}})$ & $\dots$ & $\zeta(\mathbf{t_{M,M+1}})$ & $\zeta(\mathbf{t_{M+1,M+1}}^*)$ & $\dots$ & $\zeta(\mathbf{t_{R,M+1}}^*)$\\
\hline
$\vdots$ & $\vdots$ & $\vdots$ & $\vdots$ & $\vdots$ & $\vdots$ & $\vdots$\\
\hline
$R$ & $\zeta(\mathbf{t_{1,R}})$ & $\dots$ & $\zeta(\mathbf{t_{M,R}})$ & $\zeta(\mathbf{t_{M,R}}^*)$ & $\dots$ & $\zeta(\mathbf{t_{R,R}}^*)$\\
\hline
\end{tabular}
\end{center}
}
\end{table}

\subsection{Decoding at the destination}
\label{subsec3_3}

The destination removes the CP for the first $M$ OFDM symbols and implements the following for the remaining OFDM symbols:

\begin{enumerate}
\item Remove the CP to get a $N$-point vector 
\item Shift the last $l_{cp}$ samples of the $N$-point vector as the first $l_{cp}$ samples.
\end{enumerate} 

DFT is then applied on the resulting $R$ vectors. Let the received signals for $R$ consecutive OFDM blocks after CP removal and DFT transformation be denoted by $\mathbf{y_j}=\left[\begin{array}{cccc}y_{0,j} & y_{1,j} & \dots & y_{N-1,j}\end{array}\right]^T, j=1,2,\dots,R$. Let $\mathbf{w_i}=(w_{k,i}), i=1,\dots,R$ represent the AWGN at the destination node and let $\mathbf{v_{i,j}}$ denote the DFT of $\mathbf{\bar{v}_{i,j}}$. Let $\mathbf{s_k}=\left[\begin{array}{cccc}x_{i,1} & x_{i,2} & \dots & x_{i,R}\end{array}\right]^T, k=0,1,\dots,N-1$.

Now using \eqref{eqn_identities}, we get in each sub carrier $k,0\leq k\leq N-1$: 

\begin{equation}
\label{eqn_sys_model}
\mathbf{y_{k}}=\left[\begin{array}{cccc}y_{k,1} & y_{k,2} & \dots & y_{k,R}\end{array}\right]^T=\sqrt{\frac{\pi_1\pi_2P^2}{\pi_1P+1}}\mathbf{X_kh_k}+\mathbf{n_k}
\end{equation}

\noindent where, 
\begin{equation}
\label{eqn_conj_linear}
\mathbf{X_k}=\left[\begin{array}{cccccc}\mathbf{A_1s_k} & \dots & \mathbf{A_{M}s_k} & \mathbf{A_{M+1}}\mathbf{s_k}^* & \dots \mathbf{A_{R}}\mathbf{s_k}^* \end{array}\right]
\end{equation}

\noindent for some square real matrices $\mathbf{A_i}, i=1,\dots,R$ having the property that any row of $\mathbf{A_i}$ has only one nonzero entry. If $u_k^{\tau_i}=e^{-\frac{i2\pi k\tau_i}{N}}$, then\\ 
\mbox{$\mathbf{h_k}=\left[\begin{array}{ccccccc} f_1g_1 & u_k^{\tau_2}f_2g_2 & \dots &
u_k^{\tau_{M}}f_{M}g_{M} &
u_k^{\tau_{M+1}}f_{M+1}^*g_{M+1} & \dots &
u_k^{\tau_R}f_R^*g_R\end{array}\right]^T$} is the equivalent channel matrix for the $k$-th sub carrier. The equivalent noise vector is given by

$$
\begin{array}{rl}
\mathbf{n_k}=&\sqrt{\frac{\pi_2P}{\pi_1P+1}}\left[\begin{array}{c}\delta_1\sum_{i=1}^{R}sgn(\mathbf{t_{i,1}})\mathbf{\hat{v}_{i,1}}(k)g_iu_k^{\tau_i}\\
\delta_2\sum_{i=1}^{R}sgn(\mathbf{t_{i,2}})\mathbf{\hat{v}_{i,2}}(k)g_iu_k^{\tau_i}\\
\vdots\\
\delta_R\sum_{i=1}^{R}sgn(\mathbf{t_{i,R}})\mathbf{\hat{v}_{i,R}}(k)g_iu_k^{\tau_i}
\end{array}\right]+\left[\begin{array}{c}w_{k,1}\\ w_{k,2}\\ \dots\\  w_{k,R}\end{array}\right].
\end{array}
$$

\noindent where, $sgn(\mathbf{t_{i,j}})=\left\{\begin{array}{l} 1~~\mathrm{if}~ \mathbf{t_{i,j}}\in\left\{\mathbf{r_{i,j}}, j=1,\dots,R\right\}\\
-1~~\mathrm{if}~\mathbf{t_{i,j}}\in\left\{-\mathbf{r_{i,j}}, j=1,\dots,R\right\}\\
0~~\mathrm{if}~\mathbf{t_{i,j}}=\mathbf{0}\end{array}\right.$ and\\ $\mathbf{\hat{v}_{i,m}}=\left\{\begin{array}{l} \mathbf{v_{i,j}}~\mathrm{if}~i\leq M~\mathrm{and}~\mathbf{t_{i,m}}=\pm\mathbf{r_{i,j}}\\
\mathbf{v_{i,j}}^*~\mathrm{if}~i>M~\mathrm{and}~\mathbf{t_{i,m}}=\pm\mathbf{r_{i,j}}\end{array}\right.$. The $\delta_i$'s are simply scaling factors to account for the correct noise variance due to some zeros in the transmission.

ML decoding of $\mathbf{X_k}$ can be done from \eqref{eqn_sys_model} by choosing that codeword which minimizes $\parallel\Omega^{-\frac{1}{2}}(\mathbf{y_k}-\mathbf{X_k}\mathbf{h_k})\parallel_F^2$, where $\Omega$ is the covariance matrix of $\mathbf{n_k}$ and $\parallel .\parallel_F$ denotes the Frobenius norm. Essentially, the proposed transmission scheme implements a space time code having a special structure in each sub carrier.

\subsection{Full diversity four group decodable distributed space time codes}
\label{subsec3_4}

In this subsection, we analyze the structure of the space time code required for implementing in the proposed transmission scheme. Note from \eqref{eqn_conj_linear} that the space time code should have the property that any column should have only the complex symbols or only their conjugates. We refer to this property as \textit{conjugate linearity} property\cite{RaR1,RaR2,RaR3}. But conjugate linearity alone is not enough for a space time code to qualify for implementation in the proposed transmission scheme. Note from Table \ref{table_proposed} that time reversal is implemented for certain OFDM symbol durations by all the relay nodes. In other words if one relay node implements time reversal during a particular OFDM symbol duration, then all the other relay nodes should necessarily implement time reversal during that OFDM symbol duration. Observe that this is a property connected with the row structure of a space time code. We now provide a set of sufficient conditions that are required on the row structure of conjugate linear space time codes. First let us partition the complex symbols occurring in the $i$-th row into two sets- one set $P_i$ containing those complex symbols which appear without conjugation and another set $P_i^c$ which contains those complex symbols which appear with conjugation in the $i$-th row. Then if the following conditions are satisfied by a conjugate linear space time code, then it can be implemented in the proposed transmission scheme described in the previous subsection.    

\begin{equation}
\label{eqn_row_property}
\begin{array}{c}
P_i\cap P_i^c=\phi,~\forall~i=1,\dots,R\\
|P_i|=|P_i^c|,~ \forall~ i=1,\dots,R\\
P_i\cap P_j\in\left\{\phi, P_i, P_j\right\},~\forall~ i\neq j.
\end{array}
\end{equation}

To understand what happens if the above condition is not met, let us see an example of a conjugate linear STBC which cannot be employed in the proposed transmission scheme.

\begin{eg}
Consider the conjugate linear STBC given by 
$$
\left[\begin{array}{cccc}
x_{k,1} & x_{k,2} & -x_{k,3}^* & -x_{k,4}^*\\
x_{k,2} & x_{k,3} & -x_{k,4}^* & -x_{k,1}^*\\
x_{k,3} & x_{k,4} & x_{k,1}^* & x_{k,2}^*\\
x_{k,4} & x_{k,1} & x_{k,2}^* & x_{k,3}^* 
\end{array}\right]
$$
\noindent for which $P_1=P_3^c=\left\{x_{k,1},x_{k,2}\right\}$, $P_1^c=P_3=\left\{x_{k,3},x_{k,4}\right\}$, $P_2=P_4^c=\left\{x_{k,2},x_{k,3}\right\}$, $P_2^c=P_4=\left\{x_{k,4},x_{k,1}\right\}$. It can be checked that there is no assignment of time reversal OFDM symbol durations together with an appropriate choice of $M$ and relay node processing such that the above conjugate linear STBC form is obtained at every sub carrier at the destination node. This is because the conditions in \eqref{eqn_row_property} are not met by this conjugate linear STBC.
\end{eg}

For the case of the Alamouti code, $P_1=P_2^c=\left\{x_{k,1}\right\}$, $P_2=P_1^c=\left\{x_{k,2}\right\}$ and hence it satisfies the conditions in \eqref{eqn_row_property}. Recently three new classes of full diversity four group decodable distributed space time codes for any number of relays were reported in \cite{RaR1,RaR2,RaR3}. These codes are conjugate linear. Since they are four group decodable, the associated real symbols in these space time codes can be partitioned equally into four groups and the ML decoding can be done for the real symbols in a group independently of the real symbols in the other groups. Thus the ML decoding complexity of these codes is significantly less compared to all other distributed space time codes known in the literature. In this report, we show that the codes reported in \cite{RaR1,RaR2,RaR3} satisfy the conditions in \eqref{eqn_row_property} and are thus suitable to be applied in the proposed transmission scheme. This is illustrated using the following two examples of codes taken from \cite{RaR1,RaR2,RaR3}.

\begin{eg}
\label{eg_4relay}

Let us consider $R=4$ and the distributed space time code in \cite{RaR1} for this case has the following structure
$$
\left[\begin{array}{cccc}
x_{k,1} & x_{k,2} & -x_{k,3}^* & -x_{k,4}^*\\
x_{k,2} & x_{k,1} & -x_{k,4}^* & -x_{k,3}^*\\ 
x_{k,3} & x_{k,4} & x_{k,1}^* & x_{k,2}^*\\
x_{k,4} & x_{k,3} & x_{k,2}^* & x_{k,1}^*
\end{array}\right]
$$
\noindent for which $M=2$, $P_1=P_2=P_3^c=P_4^c=\left\{x_{k,1},x_{k,2}\right\}$ and $P_3=P_4=P_1^c=P_2^c=\left\{x_{k,3},x_{k,4}\right\}$. To arrive at the above structure in every sub carrier, encoding and processing at the relays are done as follows:
$\mathbf{\bar{x}_1}=\mathrm{IDFT}(\mathbf{x_1})$, $\mathbf{\bar{x}_2}=\mathrm{IDFT}(\mathbf{x_2})$, $\mathbf{\bar{x}_3}=\mathrm{DFT}(\mathbf{x_3})$ and $\mathbf{\bar{x}_4}=\mathrm{DFT}(\mathbf{x_4})$.

\begin{table}[h]
\caption{Transmission scheme for $4$ relays}
\label{table_4relay}
\begin{center}
\begin{tabular}{|c|c|c|c|c|}
\hline
OFDM & $U_1$ & $U_2$ & $U_3$ & $U_4$\\
Symbol & & & &\\
\hline
$1$ & $\mathbf{r_{1,1}}$ & $\mathbf{r_{2,2}}$ & $-\mathbf{r_{3,3}}^*$ & $-\mathbf{r_{4,4}}^*$\\
\hline
$2$ & $\mathbf{r_{1,2}}$ & $\mathbf{r_{2,1}}$ & 
$-\mathbf{r_{3,4}}^*$ & $-\mathbf{r_{4,3}}^*$\\
\hline
$3$ & $\zeta(\mathbf{r_{1,3}})$ & $\zeta(\mathbf{r_{2,4}})$ & 
$\zeta(\mathbf{r_{3,1}}^*)$ & $\zeta(\mathbf{r_{4,2}}^*)$\\
\hline
$4$ & $\zeta(\mathbf{r_{1,4}})$ & $\zeta(\mathbf{r_{2,3}})$ & 
$-\zeta(\mathbf{r_{3,2}}^*)$ & $-\zeta(\mathbf{r_{4,1}}^*)$\\
\hline
\end{tabular}
\end{center}
\end{table}

This code is single complex symbol decodable and achieves full diversity for appropriately chosen signals sets \cite{RaR1}.
\end{eg}

\begin{eg}
\label{eg_5relay}
Let us take $R=5$ for which the distributed space time code in \cite{RaR2} is obtained taking a space time code for $6$ relays and dropping one column. It is given by 
$$
\left[\begin{array}{ccccc}
x_{k,1} & -x_{k,2}^* & 0 & 0 & 0\\
x_{k,2} & x_{k,1}^* & 0 & 0 & 0\\
0 & 0 & x_{k,3} & -x_{k,4}^* & 0\\ 
0 & 0 & x_{k,4} & x_{k,3}^* & 0\\
0 & 0 & 0 & 0 & x_{k_5}\\
0 & 0 & 0 & 0 & x_{k,6}
\end{array}\right]
$$
\noindent for which $P_1=P_2^c=\left\{x_{k,1}\right\}$, $P_2=P_1^c=\left\{x_{k,2}\right\}$, $P_3=P_4^c=\left\{x_{k,3}\right\}$, $P_4=P_3^c=\left\{x_{k,4}\right\}$, $P_5=\left\{x_{k,5}\right\}$, $P_6=\left\{x_{k,6}\right\}$ and $P_5^c=P_6^c=\phi$. At the source, we choose $\mathbf{\bar{x}_1}=\mathrm{IDFT}(\mathbf{x_1})$, $\mathbf{\bar{x}_2}=\mathrm{DFT}(\mathbf{x_2})$,
$\mathbf{\bar{x}_3}=\mathrm{IDFT}(\mathbf{x_3})$,
$\mathbf{\bar{x}_4}=\mathrm{DFT}(\mathbf{x_4})$,
$\mathbf{\bar{x}_5}=\mathrm{IDFT}(\mathbf{x_5})$ and
$\mathbf{\bar{x}_6}=\mathrm{DFT}(\mathbf{x_6})$. The $5$ relays process the received OFDM symbols as shown in Table \ref{table_5relay}.

\begin{table}[h]
\caption{Transmission scheme for $5$ relays}
\label{table_5relay}
\begin{center}
{\small
\begin{tabular}{|c|c|c|c|c|c|}
\hline
OFDM & $U_1$ & $U_2$ & $U_3$ & $U_4$ & $U_5$\\
Symbol & & & & &\\
\hline
$1$ & $\mathbf{r_{1,1}}$ & $-\mathbf{r_{2,2}}^*$ & $\mathbf{0}$ & $-\mathbf{0}$ & $\mathbf{0}$\\
\hline
$2$ & $\zeta(\mathbf{r_{1,2}})$ & $\zeta(\mathbf{r_{2,1}}^*)$ & 
$-\mathbf{0}$ & $-\mathbf{0}$ & $\mathbf{0}$\\
\hline
$3$ & $\mathbf{0}$ & $\mathbf{0}$ & $\mathbf{r_{3,3}}$ & $-\mathbf{r_{4,4}}^*$ & $\mathbf{0}$\\
\hline
$4$ & $\mathbf{0}$ & $\mathbf{0}$ & 
$-\zeta(\mathbf{r_{3,2}})$ & $\zeta(\mathbf{r_{4,1}}^*)$ & $\mathbf{0}$\\
\hline
$5$ & $\mathbf{0}$ & $\mathbf{0}$ & $\mathbf{0}$ & $\mathbf{0}$ & $\mathbf{r_{5,5}}$\\
\hline
$6$ & $\mathbf{0}$ & $\mathbf{0}$ & $\mathbf{0}$ & $\mathbf{0}$ &
$-\zeta(\mathbf{r_{5,6}})$\\
\hline
\end{tabular}
}
\end{center}
\end{table}

This code is $3$ real symbol decodable and achieves full diversity for appropriately signal sets \cite{RaR2,RaR3}.
\end{eg}

Example \ref{eg_5relay} illustrates how the proposed transmission scheme can be extended to odd number of relays as well.
 
\section{Transmission Scheme for Noncoherent\\ Asynchronous Relay Networks}
\label{sec4}

In this section, it is shown how differential encoding can be combined with the proposed transmission scheme described in Section \ref{sec3} and then the codes in \cite{RaR4} are proposed for application in this setting.

For the proposed transmission scheme in Section \ref{sec3}, at the end of one transmission frame,  we have in the $k$-th sub carrier $\mathbf{y_k}=\sqrt{\frac{\pi_1\pi_2P^2}{\pi_1P+1}}\mathbf{X_k}\mathbf{h_k}+\mathbf{n_k}$. Note that the channel matrix $\mathbf{h_k}$ depends on $f_i,g_i,\tau_i, i=1,\dots,R$. Thus the destination node needs to have the knowledge of these values in order to perform ML decoding. 

Now using differential encoding ideas which were proposed in \cite{KiR,OgH2,JiJ} for non-coherent communication in synchronous relay networks, we combine them with the proposed asynchronous transmission scheme. Supposing the channel remains approximately constant for two transmission frames, then differential encoding can be done at the source node in each sub carrier $0\leq k\leq N-1$ as follows:
$$
\mathbf{s_k^0}=\left[\begin{array}{cccc}\sqrt{R} & 0 & \dots & 0\end{array}\right]^T,~ \mathbf{s_k^t}=\frac{1}{a_t-1}\mathbf{C_t}\mathbf{s_k^{t-1}}, \mathbf{C_t}\in\mathscr{C}
$$
where, $s_k^{i}$ denotes the vector of complex symbols transmitted by the source during the $i$-th transmission frame in the $k$-th sub carrier and $\mathscr{C}$ is the codebook used by the source which consists of scaled unitary matrices \mbox{$\mathbf{C}^H\mathbf{C_t}=a_t^2\mathbf{I}$} such that $\mathrm{E}[a_t^2]=1$. If for all $\mathbf{C\in\mathscr{C}}$,

$$
\begin{array}{l}
\mathbf{C}\mathbf{A_i}=\mathbf{A_i}\mathbf{C}, i=1,\dots,M~\mathrm{and}\\ \mathbf{C}\mathbf{A_i}=\mathbf{A_i}\mathbf{C}^*, i=M+1,\dots,R 
\end{array}
$$
then we have:

\begin{equation}
\mathbf{y_k^t}=\frac{1}{a_{t-1}}\mathbf{C_t}\mathbf{y_k^{t-1}}+(\mathbf{n_k^t}-\frac{1}{a_t-1}\mathbf{C_{t}}\mathbf{n_k^{t-1}})
\end{equation}
from which $\mathbf{C_t}$ can be decoded as \mbox{$\mathbf{\hat{C}_t}=\arg\min_{\mathbf{C_t\in\mathscr{C}}}\parallel\mathbf{y_k^t}-\frac{1}{a_{t-1}}\mathbf{C_t}\mathbf{y_k^{t-1}}\parallel_F^2$} in each sub carrier $0\leq k\leq N-1$. 

Note that this decoder does not require the knowledge of $f_i,g_i,\tau_i, i=1,\dots R$ at the destination. It turns out that the four group decodable distributed differential space time codes constructed in \cite{RaR4} for synchronous relay networks with power of two number of relays meet all the requirements for use in the proposed transmission scheme as well. The following example illustrates this fact.

\begin{eg}
Let $R=4$. The codebook at the source is given by \\\mbox{$\mathscr{C}=\left\{\sqrt{\frac{1}{4}}\left[\begin{array}{cccc}
z_1 & z_2 & -z_3^* & -z_4^*\\
z_2 & z_1 & -z_4^* & -z_3^*\\
z_3 & z_4 & z_1^* & z_2^*\\
z_4 & z_3 & z_2^* & z_1^*
 \end{array}\right]\right\}$} where $\left\{z_{1I},z_{2I}\right\}$, $\left\{z_{1Q},z_{2Q}\right\}$, $\left\{z_{3I},z_{4I}\right\}$, $\left\{z_{3Q},z_{4Q}\right\}\in\mathbb{S}$ and \mbox{$\mathbb{S}=\left\{\left[\begin{array}{c}\frac{1}{\sqrt{3}}\\0\end{array}\right],\left[\begin{array}{c}-\frac{1}{\sqrt{3}}\\0\end{array}\right],\left[\begin{array}{c}0\\ \sqrt{\frac{5}{3}}\end{array}\right],\left[\begin{array}{c}0\\ -\sqrt{\frac{5}{3}}\end{array}\right] \right\}$}. Differential encoding is done at the source node for each sub carrier $0\leq k\leq N-1$ as follows:
$$
\mathbf{s_k^0}=\left[\begin{array}{cccc}\sqrt{R} & 0 & \dots & 0\end{array}\right]^T,~ \mathbf{s_k^t}=\frac{1}{a_t-1}\mathbf{C_t}\mathbf{s_k^{t-1}}, \mathbf{C_t}\in\mathscr{C}.
$$
Once we get $\mathbf{s_k^t},k=0,\dots,N-1$ from the above equation, the $N$ length vectors $\mathbf{x_i}, i=1,\dots,R$ can be obtained. Then IDFT/DFT is applied on these vectors as shown below and broadcasted to the relay nodes. $\mathbf{\bar{x}_1}=\mathrm{IDFT}(\mathbf{x_1})$, $\mathbf{\bar{x}_2}=\mathrm{IDFT}(\mathbf{x_2})$, $\mathbf{\bar{x}_3}=\mathrm{DFT}(\mathbf{x_3})$ and $\mathbf{\bar{x}_4}=\mathrm{DFT}(\mathbf{x_4})$. The relay nodes process the received OFDM symbols as given in Table \ref{table_4relay} for which $M=2$, $\mathbf{A_1}=\mathbf{I_4}$, \mbox{$\mathbf{A_2}=\left[\begin{array}{cccc}0 & 1 & 0 & 0\\
1 & 0 & 0 & 0\\
0 & 0 & 0 & 1\\
0 & 0 & 1 & 0\end{array}\right]$}, $\mathbf{A_3}=\left[\begin{array}{ccrr}0 & 0 & -1 & 0\\
0 & 0 & 0 & -1\\
1 & 0 & 0 & 0\\
0 & 1 & 0 & 0\end{array}\right]$ and \mbox{$\mathbf{A_4}=\left[\begin{array}{ccrr}0 & 0 & 0 & -1\\
0 & 0 & -1 & 0\\
0 & 1 & 0 & 0\\
1 & 0 & 0 & 0\end{array}\right]$}. It has been proved in \cite{RaR4} that $\mathbf{C}\mathbf{A_i}=\mathbf{A_i}\mathbf{C}, i=1,2$ and \mbox{$\mathbf{C}\mathbf{A_i}=\mathbf{A_i}\mathbf{C}^*, i=3,4$} for all $\mathbf{C\in\mathscr{C}}$. At the destination node, decoding for $\left\{z_{1I},z_{2I}\right\}$, $\left\{z_{1Q},z_{2Q}\right\}$, $\left\{z_{3I},z_{4I}\right\}$ and $\left\{z_{3Q},z_{4Q}\right\}$ can be done separately in every sub carrier due to the four group decodable structure of $\mathscr{C}$. 
\end{eg}

\begin{figure}[h]
\centering
\includegraphics[width=5 in]{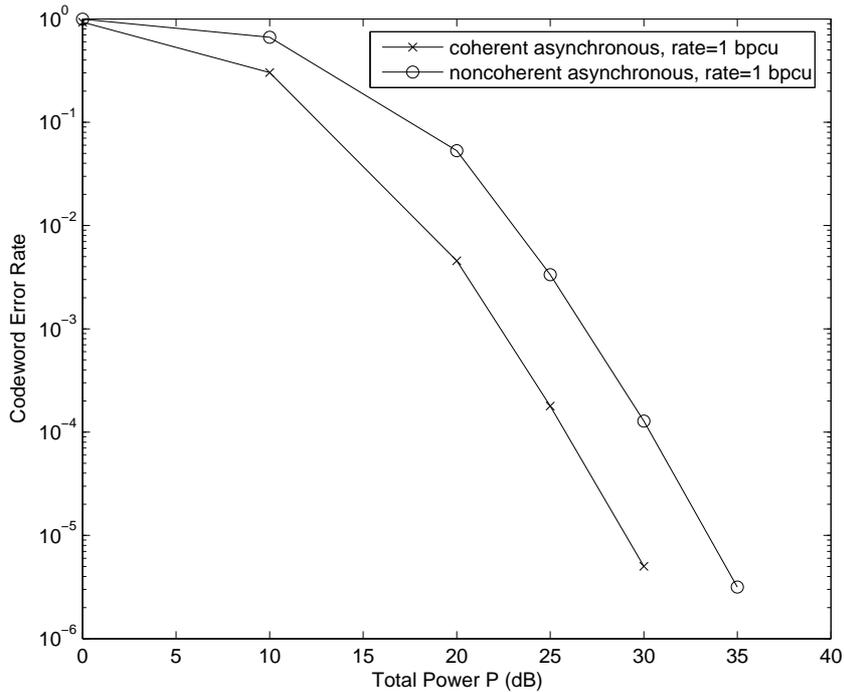}
\caption{Error performance for a $4$ relay system with and without channel knowledge}
\label{fig_simulation}
\end{figure}

\section{Simulation results}
\label{sec5}
In this section, we study the error performance of the proposed codes using simulations. We take $R=4$, $N=64$ and the length of CP as $16$. The delay $\tau_i$ at each relay is chosen randomly between $0$ to $15$ with uniform distribution. Two cases are considered for simulation: (1) with channel knowledge at the destination and (2) without channel knowledge at the destination. For the case of no channel information, differential encoding at the source as described in Section \ref{sec4} is done using the distributed differential space time in \cite{RaR4}. When channel knowledge is available at the destination, rotated QPSK is used as the signal set \cite{RaR1,RaR3}. The transmission rate for the both the schemes is $1$ bit per channel use (bpcu) if the rate loss due to CP is neglected. 

The error performance curves for both the cases is shown in Fig. \ref{fig_simulation}. It can be observed from Fig. \ref{fig_simulation} that the error performance of the no channel knowledge case performs approximately $5$ dB worser than that with channel knowledge at the destination. This is due to the differential transmission/reception technique in part and also in part because of the change in signal set from rotated QPSK to some other signal set \cite{RaR4} in order to comply with the requirement of scaled unitary codeword matrices. The change in signal set for the sake of scaled unitary codeword matrices results in a reduction of the coding gain.

\section{Discussion}
\label{sec6}

A general transmission scheme for arbitrary number of relays that can achieve full cooperative diversity in the presence of timing errors at the relay nodes was proposed. It was then pointed out that the four group decodable distributed space time codes in \cite{RaR1,RaR2,RaR3} can be applied in the proposed transmission scheme for any number of relay nodes. Finally it was shown how the proposed scheme can be combined with differential encoding at the source node to end up with a transmission scheme that is robust to timing errors and also does not require the knowledge of the channel fading gains as well as the timing errors at any of the nodes. For this differential scheme, it was pointed out that the four group decodable distributed differential space time codes in \cite{RaR4} are applicable for power of two number of relays. 

A drawback of the proposed transmission scheme is that it requires a large coherence interval spanning over multiple OFDM symbol durations. Moreover there is a rate loss due to the use of CP, but this loss can be made negligible by choosing a large enough $N$. Some of the interesting directions for further work are listed below:

\begin{enumerate}
\item Constructing single symbol decodable distributed space time codes for the proposed transmission scheme. 
\item The codes in \cite{RaR4} are applicable only for power of two number of relay nodes. Constructing four group decodable distributed differential space time codes for all even number of relay nodes that are applicable in asynchronous relay networks without channel knowledge is an important direction for further work.
\item In this work, we have assumed that there are no frequency offsets at the relay nodes. Extending this work to asynchronous relay networks with frequency offsets is an interesting direction for further work. This problem has been addressed in \cite{LiX2} for the case of two relay nodes. 
\end{enumerate}

\section*{Acknowledgement}
The authors sincerely thank Prof. Xiang Gen Xia and Prof. Hamid Jafarkhani for sending us preprints of their recent works \cite{LiX,GuX,JiJ,LiX2}.


                                                                                                                                                             
\end{document}